\documentclass[twocolumn,amsmath,amssymb,prb,superscriptaddress,fleqn]{revtex4-1}
\usepackage{epsfig}
\usepackage{graphicx}
\usepackage{dcolumn}
\usepackage{bm}
\usepackage{color} 
\newcommand{\fs}[1]{\textcolor{black}{#1}}
\usepackage{ulem}
\usepackage{amsmath}
\usepackage{amsthm}

\usepackage{multirow}
\usepackage{amsthm}

\newtheoremstyle{query}%
{}{}
{\color{red}}
{}
{\sffamily\bfseries}{:}{12pt}
{}
\theoremstyle{query}

\begin{document}


\title{Scaling the electrical current switching of exchange bias in fully-epitaxial antiferromagnet/ferromagnet bilayers}

\author{T. Hajiri}
\altaffiliation[Electronic mail:~]{t.hajiri@nagoya-u.jp}
\affiliation{Department of Materials Physics, Nagoya University, Nagoya 464-8603, Japan}
\author{H. Goto}
\affiliation{Department of Materials Physics, Nagoya University, Nagoya 464-8603, Japan}
\author{H. Asano}
\affiliation{Department of Materials Physics, Nagoya University, Nagoya 464-8603, Japan}
\date{\today}
%
\begin{abstract}
While the electrical current manipulation of antiferromagnets (AFMs) has been demonstrated, the extent of the studied AFM materials has been limited with few systematic experiments and a poor understanding.
We compare the electrical current switching of the exchange-bias field ($H_{ex}$) in AFM-Mn$_3A$N/ferromagnet-Co$_3$FeN bilayers.
An applied pulse current can manipulate $H_{ex}$ with respect to the current density and \fs{FM layer magnetization}, which shifts exponentially as a function of the current density.
We found that the saturation current density and exponential decay constant $\tau$ \fs{increase with} the local moment of AFM Mn atoms.
Our results highlight the effect of the AFM local moment to electrical current switching of $H_{ex}$, although it has a near-zero net magnetization, and may provide a facile way to explore the electrical current manipulation of AFM materials.
\end{abstract}


\maketitle
\section{INTRODUCTION}
Electrical manipulation of magnetic moments of ferromagnets (FMs) using spin-transfer torque (STT) and spin-orbit torque (SOT) has been widely studied towards spintronic applications~\cite{STT1, STT2, SOT1, SOT2}.
Recently, antiferromagnetically coupled materials, such as ferrimagnets, synthetic antiferromagnets (AFMs), and AFMs, have attracted significant attention due to their promising potential in spintronic applications because of several properties, such as the absence of stray magnetic fields, terahertz spin dynamics, and stability against external perturbations~\cite{AFM1, AFM2, AFM3}.
The electrical current switching of the AFM N$\rm{\acute{e}}$el vector has been demonstrated for several AFM materials, such as CuMnAs~\cite{SOT_CuMnAs1, SOT_CuMnAs2}, Mn$_2$Au~\cite{SOT_Mn2Au1, SOT_Mn2Au2}, NiO~\cite{SOT_NiO1, SOT_NiO2, SOT_NiO3}, and Fe$_2$O$_3$~\cite{SOT_Fe2O3}, with electrical current densities on the order of 10$^7$–-10$^8$A/cm$^2$, and recently switching with the order of $10^6$~A/cm$^2$ has been realized on noncollinear AFM of Mn$_3$GaN~\cite{SOT_MGN} and Mn$_3$Sn~\cite{SOT_Mn3Sn}.
In addition, the electrical current switching of ferrimagnets has been demonstrated.
In particular, the magnetic moments of transition-metal (TM) and rare-earth (RE) ferrimagnets can be aligned in an antiparallel orientation, which enables tuning the magnetization around the compensation points by varying the relative constituent concentrations.
The torque efficiency increases with divergent behaviors due to negative exchange interactions towards the compensation points~\cite{RE-TM1, RE-TM2}.
On the other hand, in AFM, there are few systematical studies due to the difficulty of reading out the AFM N$\rm{\acute{e}}$el vector electrically as only a few AFM materials exhibit large electrical signals as an anomalous Hall effect~\cite{AHE1, AHE2, Cu-MNN_AHE} and anisotropic magnetoresistance~\cite{TAMR_NMat, TAMR_PRL, FCS_RMG, TAMR_PRL2}.

To address this issue, we focused on the electrical switching of the exchange-bias field $H_{ex}$, which was first reported more than a decade ago~\cite{AFM_STT_PRL, AFM_STT_APL}.
The exchange-bias effect has been widely studied because of its general use in spintronic devices, such as for spin-valve type magnetic memory~\cite{MRAM1, MRAM2}.
In general, the exchange-bias effect occurs at the interfaces between FM and AFM materials, regardless of the material combination~\cite{Radu and Zabel}.
Therefore, in principle, electrical switching for $H_{ex}$ can be adapted to all AFM materials.
The early studies of electrical switching for $H_{ex}$ demonstrated that spin torque can act not only on FM but also on AFM order parameters.
Several related studies were conducted to further explore this observation~\cite{AFM_STT1, AFM_STT2, AFM_STT3, Hex_SOT, Hex_EDL}; however, no clear relationship was reported between the $H_{ex}$ switching behavior and the electrical current density.

Based on theoretical studies of the electrical current switching of the AFM N$\rm{\acute{e}}$el vector in AFM/FM bilayers where the spin-polarized current is injected into the AFM through the FM layer, the critical current density $i_c$ is expressed as
\begin{eqnarray}
\begin{aligned}
i_{c, AFM}&=&\frac{2\gamma_{AFM}\omega_{AFM}M_{AFM}d_{AFM}}{\gamma\sigma H_{flop}}\\
&\propto&\frac{\gamma_{AFM}}{\omega_{AFM}}H^{AFM}_{an}, \qquad\sigma\equiv\frac{\mu_B}{e}\varepsilon P,
\label{eq:one}
\end{aligned}
\end{eqnarray}
where $\gamma_{AFM}$ is the coefficient of internal friction, $\omega_{AFM}$ is the antiferromagnetic resonance frequency, $H_{flop}$ is the flop field, $M_{AFM}$ is the absolute value of the AFM local moment, $d_{AFM}$ is the thickness of the AFM layer, $\gamma$ is the absolute value of the gyromagnetic ratio, $H^{AFM}_{an}$ is the anisotropy field of AFM, $\mu_B$ is the Bohr magneton, $P$ is the degree of spin polarization, $e$ is the electron charge, and $\varepsilon$ is the efficiency of the spin polarization ($\leq1$)~\cite{Gomonay1, Gomonay2}.
For FM, $i_{c,FM}$ can be expressed as $i_{c,FM}\propto(\gamma_{AFM}/\omega_{AFM})H^{FM}_{an}$.
In these theory, they assume $H^{AFM}_{an}$ of $\sim0.1$~kOe (e.g., FeMn, IrMn, NiO), and those values are lower than in typical FM, so $i_{c, AFM}$ is expected to be smaller than $i_{c, FM}$~\cite{Gomonay1, Gomonay2}.
Our previous studies demonstrated that the electrical switching of $H_{ex}$ is realized in antiperovskite nitride AFM-Mn$_3$GaN/FM-Co$_3$FeN (001) bilayers with a current density of $1\times10^6$A/cm$^2$, which is two orders of magnitude smaller than typical FM cases~\cite{Sakakibara_bilayer}.
Towards further understanding, we introduce that the noncollinear AFM antiperovskite manganese nitride Mn$_3A$N may be a good platform because the AFM magnetic properties, such as the Mn local moment and AFM spin structure ($\Gamma_{4g}$ and $\Gamma_{5g}$), depend on the $A$ atom~\cite{MAN1, MAN2, MAN3}.

In this study, we prepared fully epitaxial AFM-Mn$_3A$N/FM-Co$_3$FeN (001) bilayers with $A$~=~Ga, Ni, and Ni$_{0.35}$Cu$_{0.65}$ (hereinafter referred to as MGN, MNN, and Cu-MNN), which have similar magnetic properties.
Shifts in $H_{ex}$ as a function of current density show an exponential change above the threshold current density, regardless of the AFM Mn$_3A$N layer.
We clearly show that the electrical current switching of $H_{ex}$ depends on the AFM local moment of the M$A$N, even though it has a zero net magnetization.
These results highlight that the spin-polarized current has a different degree of action on the AFM local moment depending on its magnitude.

\begin{figure}[b]
\includegraphics[width=\linewidth,clip]{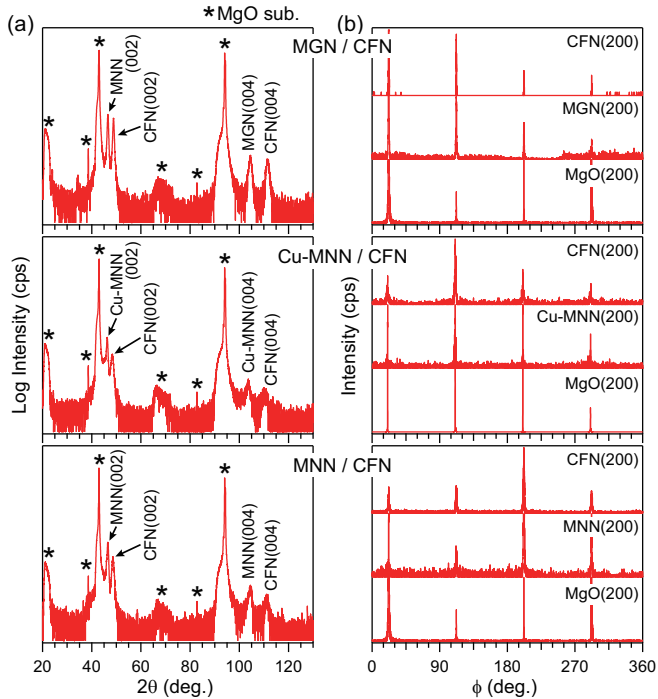}
\caption{
(a) Out-of-plane and (b) in-plane $\phi$-scans of the considered M$A$N/CFN (001) bilayers.
}
\label{fig:one}
\end{figure}

\section{EXPERIMENTAL DETAILS}
The Mn$_3A$N (20~nm, bottom)/Co$_3$FeN (10~nm, top) bilayers (hereinafter referred to as M$A$N and CFN) were grown using reactive magnetron sputtering on MgO(001) substrates under a total pressure of 2.0 Pa for an Ar + N$_2$ gas mixture.
Details of the film and bilayer growth were reported in our previous works~\cite{Sakakibara_CFN, Sakakibara_bilayer, Cu-MNN_AHE, Cu-MNN_growth}.
The crystal structure was analyzed using both in-plane and out-of-plane X-ray diffraction (XRD) measurements with Cu~$K\alpha$ radiation. 
The magnetic properties of the M$A$N/CFN bilayers were characterized using superconducting quantum interference device (SQUID) magnetometry, and the transport properties were characterized with the standard DC four-terminal method in the current-in-plane configuration with $20~\mu$m-wide Hall bars prepared using conventional photolithographic processes. 
To induce exchange coupling between the FM and AFM layers, the M$A$N/CFN bilayers were cooled from 350~K to 4~K under an applied magnetic field of +10~kOe along the M$A$N/CFN~[110] direction.

\begin{figure}[b]
\includegraphics[width=\linewidth,clip]{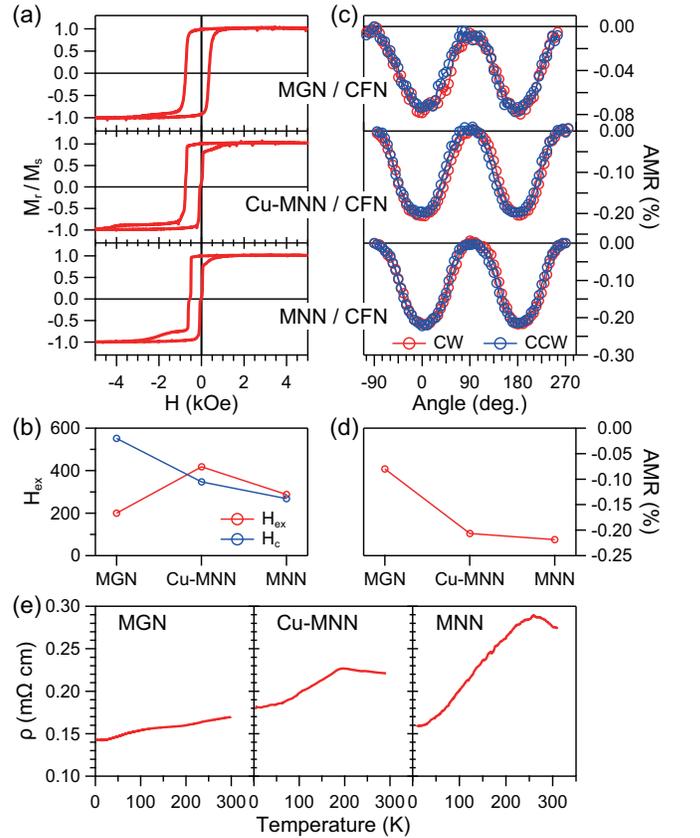}
\caption{(a) Magnetic hysteresis loops for the M$A$N/CFN bilayers at $T=4$~K after FC.
The measurements were performed along the easy axis of CFN $\langle110\rangle$.
(b) $H_c$ and $H_{ex}$ as functions of the M$A$N layer.
(c) AMR properties of M$A$N/CFN bilayers, where 0$^\circ$ is the sensing current direction parallel to the M$A$N/CFN[110] direction.
The magnetic field of 5~kOe is rotated in the film plane.
(d) AMR ratio as a function of the M$A$N layer.
(e) Temperature-dependent resistivity of 50~nm thick M$A$N single-layer films.
}
\label{fig:two}
\end{figure}

\begin{figure*}[t]
\includegraphics[width=\linewidth,clip]{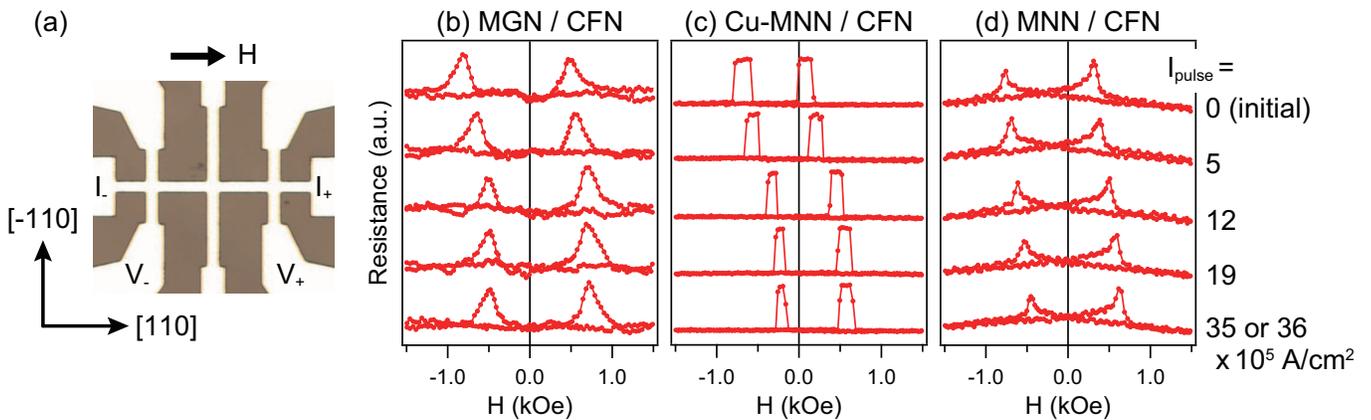}
\caption{
(a) Microscope image of M$A$N/CFN bilayers and experimental configuration.
(b--d) AMR curves after applying several 100~msec width pulse currents ($I_{pulse}$) into the M$A$N/CFN bilayers.
All measurements were performed at $T=4$~K.
}
\label{fig:three}
\end{figure*}

\section{RESULTS AND DISCUSSIONS}
The out-of-plane and in-plane XRD patterns of the M$A$N/CFN bilayers are shown in Figs.~\ref{fig:one}(a) and~\ref{fig:one}(b), respectively.
In Fig.~\ref{fig:one}(a), only the $(00l)$ M$A$N and $(00l)$ CFN series exhibit Bragg peaks in the out-of-plane XRD patterns, which indicates the (001)-oriented M$A$N/CFN bilayers grow well on MgO(001) substrates. 
In addition, epitaxial growth is confirmed in all M$A$N/CFN bilayers from the results of the $\phi$-scan measurement, as shown in Fig.~\ref{fig:one}(b), showing that their epitaxial relationship is MgO(001)[100]//M$A$N(001)[100]//CFN(001)[100].

Figure~\ref{fig:two}(a) shows the magnetic hysteresis loops for M$A$N/CFN bilayers measured along the $\langle110\rangle$ direction at $T=4$~K after field cooling (FC).
All loops exhibit a clear hysteresis loop shift, indicating the exchange-bias effect appears in all M$A$N/CFN bilayers.
The M$A$N-layer dependence of the coercive field ($H_c$) and the exchange-bias field ($H_{ex}$) are illustrated in Fig.~\ref{fig:two}(b).
The M$A$N layer dependence of the anisotropic magnetoresistance (AMR) under an applied magnetic field of 5~kOe after FC is presented in Fig.~\ref{fig:two}(c).
The AMR ratio is defined as $\rm{AMR}=\{\rho(\theta)-\rho(90^\circ)\}/\rho(90^\circ)$, where $\theta$ is the angle between the sensing current parallel to the [110] direction and applied magnetic field in the (001) film plane.
Each AMR curve exhibits negative values due to the high polarization of the CFN layer~\cite{Sakakibara_CFN, takahashi_CFN, Sakuraba_AMR} with no dependence on the rotational direction.
In the single-layer CFN films and MGN/CFN bilayers, the AMR ratios are reported to be approximately -0.8\% and -0.09\%, respectively~\cite{Sakakibara_CFN, Sakakibara_bilayer}.

As summarized in Fig.~\ref{fig:two}(d), the AMR ratios are similar at approximately -0.1 -- -0.2\% to the previous MGN/CFN bilayers.
Figure~\ref{fig:two}(e) shows the temperature-dependent resistivity of M$A$N single-layer films.
The N$\rm{\acute{e}}$el temperatures of Cu-MNN and MNN are observed at 190~K and 250~K, while that of MGN is higher than 300~K, of which values are consistent from previous report~\cite{Cu-MNN_AHE, Cu-MNN_growth, MGN_TN1, MGN_TN2}.
Although $H_c$ and $H_{ex}$ show small variations by the M$A$N layer, no significant exchange-spring effects, such as an enhanced $H_c$ and rotation-direction dependence of AMR~\cite{TAMR_NMat, TAMR_PRL, FCS_RMG}, are observed in each M$A$N/CFN biayers.
Besides, the resistivity of each M$A$N single-layer film at $T=4$~K is similar at approximately 0.14 -- 0.18~m$\Omega$cm.
Thus, these results suggest that each exchange coupling can be compared in electrical switching.

In the electrical current switching for the $H_{ex}$ in M$A$N/CFN bilayers, Fig.~\ref{fig:three}(a) shows a microscope image of the M$A$N/CFN bilayers and experimental configuration.
During FC, a magnetic field of 10.0~kOe was applied along the [110] direction.
When the pulse current was injected, a magnetic field of $H_{switch}=-5.0$~kOe was applied to define the \fs{CFN layer magnetization}, which is the opposite of the FC direction.
Before the initial magnetoresistance (MR) measurements after FC was performed, the magnetic field was swept between $\pm5.5$~kOe several times to avoid the training effect~\cite{training_effect}.
The MR curves after applying several pulse currents ($I_{pulse}$) with 100~ms width into the M$A$N/CFN bilayers are shown in Figs.~\ref{fig:three}(b--d).
Here, the current density is calculated from total thickness of each bilayers.
In the initial MR curves, the $H_c$ and $H_{ex}$ obtained from the two MR peaks agree well with those measured using SQUID, as shown in Fig.~\ref{fig:two}(a), which allows estimating $H_{ex}$ switching from the MR measurements.
After applying $I_{pulse}=5\times10^5$~A/cm$^2$, the MR peaks shift towards the positive field direction for all M$A$N/CFN bilayers, indicating that $H_{ex}$ shifts in the opposite direction to the initial $H_{ex}$ as induced from the FC.
By increasing $I_{pulse}$, the shift in $H_{ex}$ gradually increases but with a different saturation behavior.
For the MGN/CFN bilayers, the shift in $H_{ex}$ saturates between $5-12 \times10^5$~A/cm$^2$, while that of Cu-MNN/CFN saturates between $12-19 \times10^5$~A/cm$^2$.
On the other hand, those of the MNN/CFN bilayers continue to shift in the presented MR curves for the $I_{pulse}$ range.
We note that the sensing current was set to approximately $5\times10^3$~A/cm$^2$, which is sufficiently small for $H_{ex}$ switching.
The observed $H_{ex}$ switching is reversible with respect to \fs{FM layer magnetization during $I_{pulse}$ injection}, as shown in Fig.~S1~\cite{Supplement} and discussed later.

\begin{figure}[t]
\includegraphics[width=\linewidth,clip]{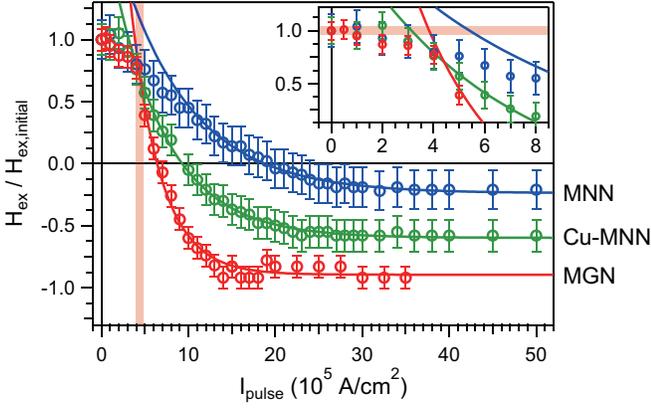}
\caption{
Shifts of $H_{ex}$ normalized by $H_{ex, ini.}$ as a function of $I_{pulse}$.
The inset shows an enlarged image around $I_{pulse}\sim8\times10^5$~A/cm$^2$.
The bold solid lines are the results of a fit to the exponential decay function $y=y_0+Ae^{-(x+x_0)/\tau}$.
}
\label{fig:four}
\end{figure}

\begin{figure}[t]
\includegraphics[width=0.9\linewidth,clip]{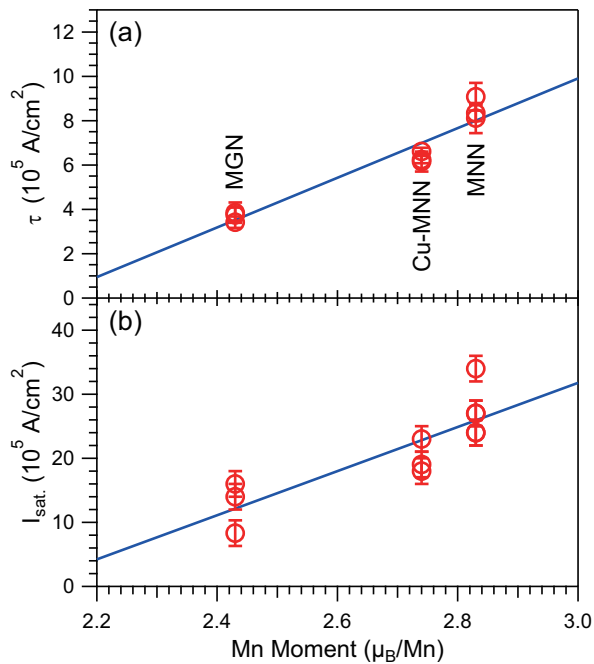}
\caption{
(a) Decay constants $\tau$ and (b) $I_{sat.}$ as functions of the Mn moment for Mn$_3A$N.
The bold solid lines are linear fits.
}
\label{fig:nine}
\end{figure}

Figure~\ref{fig:four} presents the shifts of $H_{ex}$ normalized by the initial exchange-bias field $H_{ex, ini.}$ as a function of $I_{pulse}$.
The shift begins from approximately $I_{pulse}=1\sim3\times10^5$~A/cm$^2$, which is commonly observed.
On the other hand, applying an additional $I_{pulse}$ produces different shift behaviors.
For MGN, the $H_{ex}/H_{ex, ini}$ curve becomes steeper above $4\times10^5$~A/cm$^2$ and saturates at $H_{ex}/H_{ex, ini}=-0.9$ with a saturation current density of $I_{sat.}$ of $12\times10^5$~A/cm$^2$.
For Cu-MNN, the $H_{ex}/H_{ex, ini}$ curve analogously becomes steeper at $4\times10^5$~A/cm$^2$, but with a smaller slope.
The $H_{ex}/H_{ex, ini}$ saturates at $-0.55$ around $22\times10^5$~A/cm$^2$.
On the other hand, the MNN linearly changes up to $12\times10^5$~A/cm$^2$ and saturates at $-0.25$ around $28\times10^5$~A/cm$^2$.
Above 4 or $12\times10^5$~A/cm$^2$, the shift of $H_{ex}$ can be fit using the exponential decay function $y=y_0+Ae^{-(x+x_0)/\tau}$.
The decay constants $\tau$ for MGN, Cu-MNN, and MNN are 3.42, 6.25, and 8.11\fs{$\times10^5$~A/cm$^2$}, respectively. 
These results indicate that the effects of spin-polarized current on the exchange bias are different between the considered M$A$N layers.
We note that the saturation $H_{ex}/H_{ex, ini}$ depends on the samples, regardless of the M$A$N layer while the $\tau$ does not show a sample dependence, as illustrated in Fig.~\ref{fig:nine}.
Generally, the exchange-bias effect is strongly affected by the interface frustration, roughness and other properties~\cite{Reorientation_poly, interface_roughness}, which may affect the saturation $H_{ex}/H_{ex, ini}$ and requires future study.

\begin{figure}[t]
\includegraphics[width=\linewidth,clip]{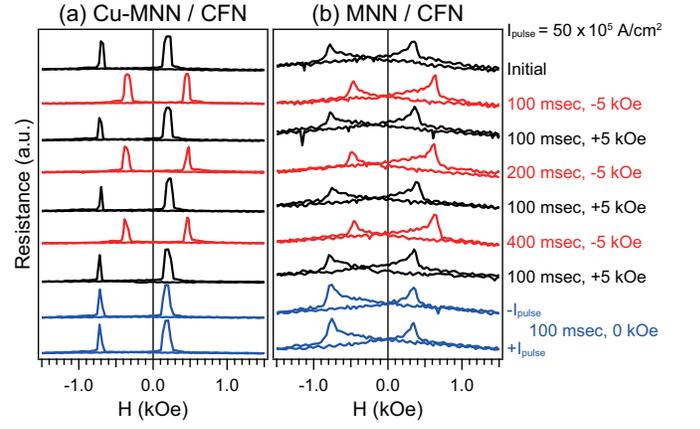}
\caption{
Sequential switching and reversing results of (a) Cu-MNN/CFN and (b) MNN/CFN bilayers, respectively.
}
\label{fig:five}
\end{figure}

Although it is considered that the torques from the spin-polarized current play an important role in $H_{ex}$ switching as reported in the previous study~\cite{AFM_STT_PRL, AFM_STT_APL, AFM_STT1, AFM_STT2, AFM_STT3}, to exclude other possible origin, several sequential measurements were performed in Cu-MNN/CFN and MNN/CFN bilayers as shown in Fig.~\ref{fig:five}(a) and \ref{fig:five}(b), respectively.
After the initial MR measurements, $I_{pulse}=50\times10^5$~A/cm$^2$ is injected with 100~msec pulse under -5~kOe (switching process).
As discussed in Fig.~\ref{fig:three}, the MR peaks shift towards the positive field direction.
Then, $I_{pulse}$ is injected under +5.0 kOe (reversing process), and the MR curves return to the initial state with the initial $H_{ex}$.
These results indicate that opposite polarity of $H_{ex}$ also can be manipulated and \fs{the CFN layer magnetization} with respect to $H_{ex}$ direction plays an important role\fs{, probably due to injection of spin-polarized current from CFN layer to M$A$N layer}.
In the same manner, switching and reversing processes were sequentially performed with 200 and 400~msec pulse under -5~kOe for switching and 100~msec pulse under +5~kOe for reversing.
Regardless of the switching pulse width, the $H_{ex}$ after switching shows the same value, and similarly, the $H_{ex}$ after reversing shows the same value as initial $H_{ex}$, suggesting no sizable difference of thermal effect in these pulse-width range.
In addition, $I_{pulse}=\pm50\times10^5$~A/cm$^2$ was injected with 100~msec pulse without magnetic field.
If significant heating and subsequent cooling process are occurred by $I_{pulse}$, the exchange-coupling properties are expected to be affected when there is no magnetic field.
However, no sizable difference by applying $\pm I_{pulse}$ is observed in not only the magnitude of $H_{ex}$ but also shape of MR curves in both Cu-MNN/CFN and MNN/CFN bilayers.
\fs{Besides, the temperature increase during 100~msec pulse duration was measured as shown in supplemental Fig. S2~\cite{Supplement}.
The temperature reached until the $H_{ex}$ switching is completed is sufficiently lower than the N$\rm{\acute{e}}$el temperature.}
From these results, we conclude that most switching effect comes from spin-polarized current rather than thermal effect.

\begin{figure}[b]
\includegraphics[width=\linewidth,clip]{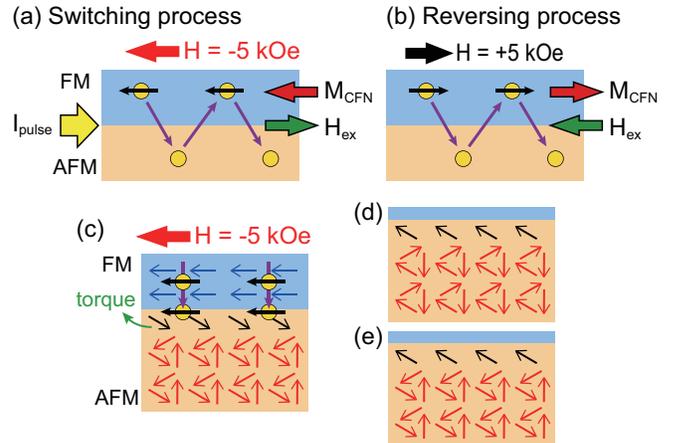}
\caption{
Schematic illustrations of possible mechanism of $H_{ex}$ switching. 
(a) Switching process under -5~kOe and (b) reversing process under +5~kOe.
(c) Enlarged schematic illustration at the interface between FM and AFM.
When the current flows from FM to AFM, the current is spin polarized and the spin-polarized current acts (f) only interface uncompensated spins or (g) not only interface uncompensated spins, but also to a certain depth of bulk AFM spins.
}
\label{fig:six}
\end{figure}

The schematic illustration of possible $H_{ex}$ switching mechanism are presented in Fig.~\ref{fig:six}.
Since the resistivity of CFN is similar at around 0.1~m$\Omega$cm with each M$A$N layer, here we assume that current can flow back and forth between FM and AFM as illustrated in Fig.~\ref{fig:six}(a) and \ref{fig:six}(b). 
When the current flows from CFN to M$A$N, the current is spin polarized due to CFN spin polarization and its polarized direction can be directed in the opposite direction to $H_{ex}$ by the magnetic field as illustrated in Fig.~\ref{fig:six}(a)--\ref{fig:six}(c).
The previous study argued that the spin-polarized current induces torques on interface uncompensated spins as illustrated in Fig.~\ref{fig:six}(d).
On the other hand, resent SOT studies of noncollinear AFM Mn$_3$Sn demonstrate the manipulation of AFM moments even at 40~nm thick films~\cite{SOT_Mn3Sn}.
Hence, as illustrated in Fig.~\ref{fig:six}(e), the spin polarized current may be able to act not only on interface uncompensated spins but also to a certain depth of bulk AFM spins.
Although it is difficult to distinguish these two possible pictures in this study, in the below discussion, we assume that interface AFM spins which cause $H_{ex}$ keep the bulk AFM properties, such as the local moment of the AFM magnetic atoms, the anisotropy of the AFM, and the AFM N$\rm{\acute{e}}$el vector.

\begin{figure}[t]
\includegraphics[width=\linewidth,clip]{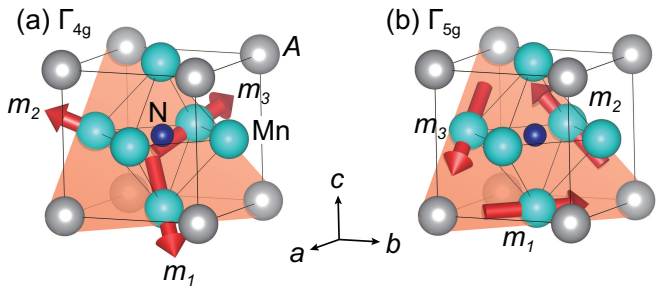}
\caption{
(a) $\Gamma_{4g}$ and (b) $\Gamma_{5g}$ AFM spin structures visualized using VESTA package~\cite{VESTA}.
}
\label{fig:seven}
\end{figure}
\begin{table}[b]
\begin{center}
\caption{
Relative angles between the AFM N$\rm{\acute{e}}$el vectors (${\bm L}_1$, ${\bm L}_2$) and the $\langle110\rangle$ direction for $\Gamma_{4g}$ and $\Gamma_{5g}$ AFM orders.
}
\begin{tabular}{{>{\centering\arraybackslash}p{1.9cm}}*{4}{>{\centering\arraybackslash}p{1.5cm}}}
\hline \hline
AFM order & & ${\bm L}_1$ & ${\bm L}_2$ & Average \\ \hline
\multirow{2}{*}{$\Gamma_{4g}$} & $\theta$ & 73$^\circ$ & 60$^\circ$ & --\\
& sin$\theta$ & 0.96 & 0.87 & 0.91\\ \hline
\multirow{2}{*}{$\Gamma_{5g}$} & $\theta$ & 60$^\circ$ & 107$^\circ$ & --\\
& sin$\theta$ & 0.87 & 0.96 & 0.91\\
\hline \hline
\end{tabular} 
\label{table:one}
\end{center}
\end{table}

Finally, we would like to discuss the different $H_{ex}$ switching behaviors.
According to theoretical studies of STT in AFM materials, the critical current is affected by the local moment of the AFM magnetic atoms, the anisotropy of the AFM, and the relative angle between the injected spin direction and the AFM N$\rm{\acute{e}}$el vector~\cite{Gomonay1, Gomonay2, Nogues}.
In M$A$N systems, the AFM spin orders ($\Gamma_{4g}$ and $\Gamma_{5g}$, shown in Fig.~\ref{fig:seven}) vary depending on the composition of the $A$ site; $\Gamma_{5g}$ is observed in MGN and MNN and $\Gamma_{4g}$ in Cu-MNN.
In such noncollinear AFMs with triangular magnetic configurations composed of three equivalent magnetic directions (${\bm m_1}$, ${\bm m_2}$, and ${\bm m_3}$, shown in Fig.~\ref{fig:seven}), the AFM N$\rm{\acute{e}}$el vectors can be expressed using two vectors ${\bm L}_1$ and ${\bm L}_2$ as~\cite{Gomonay_3},
\begin{eqnarray}
\begin{aligned}
{\bm L}_1&=&{\bm m}_1+{\bm m}_2-2{\bm m}_3\\
{\bm L}_2&=&\sqrt{3}({\bm m}_1-{\bm m}_2).
\label{eq:one}
\end{aligned}
\end{eqnarray}
A theoretic study indicates that the torque efficiency is proportional to sin$\theta$, where $\theta$ is the relative angle between the injected spin direction and the AFM N$\rm{\acute{e}}$el vector~\cite{Nogues}.
In this study, the injected spin direction from the FM CFN layer is $\langle110\rangle$ as determined from the external magnetic field.
The relative angles between ${\bm L}_1$, ${\bm L}_2$ and $\langle110\rangle$ are summarized in Table.~\ref{table:one}.
From the perspective of sin$\theta$, there is no defference between $\Gamma_{4g}$ and $\Gamma_{5g}$ orders.
Therefore, we conclude that the effect of the relative angle between the injected spin direction and the AFM N$\rm{\acute{e}}$el vector is relatively small.

\begin{figure}[t]
\includegraphics[width=0.6\linewidth,clip]{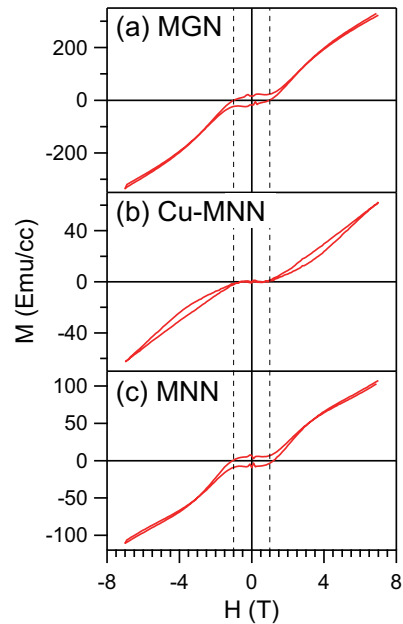}
\caption{
In-plane magnetic hysteresis loops for (a) MGN, (b) Cu-MNN , and (c) MNN (111) films measured at $T=4$~K.
The magnetic field is applied along the [-1-12] direction.
}
\label{fig:eight}
\end{figure}

On the other hand, although there are no reports on the actual magnetocrystalline anisotropy in M$A$N systems, it is reported that the Mn moments of Cu-MNN begin to cant in the (111) plane when a magnetic field is applied along the [11-2] direction with a magnitude that is larger than the threshold value~\cite{Cu-MNN_growth}.
Thus, epitaxial-grown M$A$N(111) films are prepared~\cite{Cu-MNN_AHE, Cu-MNN_growth} and compared the magnetization loop along the [-1-12] direction for each M$A$N(111) film as shown in Fig.~\ref{fig:eight}.
All M$A$N (111) films show the similar in-plane magnetization behavior with a threshold value of 1~T, indicating that these AFM materials have similar anisotropies, at least for canting in the (111) plane.
Therefore, it is considered that the effects of both the relative angle between the injected spin direction and the AFM N$\rm{\acute{e}}$el vector and the anisotropy are small.

In M$A$N systems, the Mn local moment is known to depend on the $A$ atom and is proportional to the number of electrons outside of the Ar closed shells~\cite{MAN1}.
From theoretical calculations~\cite{MAN2}, the Mn local moments for MGN and MNN are 2.43 and 2.83~$\mu_B$/Mn atom, respectively, and that of Cu-MNN is estimated as 2.74~$\mu_B$/Mn atom.
Figure~\ref{fig:nine} compares the decay constant $\tau$ and saturation current $I_{sat.}$ as a functions of the Mn local moments for M$A$N, indicating that both the $\tau$ and $I_{sat.}$ \fs{increase with} Mn local moments of M$A$N.
Although $I_{sat.}$ shows certain sample dependence probably due to the interface condition, even taking into account those variations, the estimated linear relationship \fs{seems to be} valid.
This conclusion can be reinforced by $\tau$ because it shows a small sample dependence.

In addition, to check the universality of this relationship, we considered collinear AFM MnN/CFN bilayers, because the MnN has a rocksalt structure with a much higher N$\rm{\acute{e}}$el temperature at 650~K and larger AFM local moment of 3.3~$\mu_B$/Mn than for M$A$N~\cite{MnN, MnN_MOKE1, MnN_MOKE2}.
As shown in \fs{Fig.~S3}~\cite{Supplement}, the $H_{ex}/H_{ex,ini}$ shifts with respect to $I_{pulse}$ in the same manner, and $I_{sat.}$ as a function of the Mn local moment follows an almost proportional relationship.
These results indicate that the spin-polarized current has a different degree of action on the AFM local moment depending on its magnitude.
The linear fitting line intersects the $x$-axis in the vicinity of 2~$\mu_B$/Mn.
Since a scaling is not expected to be achieved for the local moments of less than $2\mu_B$/atom, switching experiments with such AFMs are of interest.
Typical AFMs used in spintronics such as MnIr and MnPt, however, exhibit large local magnetic moments of $3\sim4\mu_B$/Mn.
Another possibility would be a double helical AFM MnP~\cite{MnP} and CrAs~\cite{CrAs}, which have $1.3\mu_B$/Mn and $1.7\mu_B$/Cr, respectively, and it is worthy of future research on those AFM switching.
We note that the similar order of switching current density has been recently realized in the SOT of noncollinear AFM of MGN(2.43~$\mu_B$/Mn) and Mn$_3$Sn(3.2~$\mu_B$/Mn~\cite{Mn3Sn_moment}) with 1.5 and $5\times10^6$~A/cm$^2$, respectively~\cite{SOT_MGN, SOT_Mn3Sn}.
Since STT consumes more energy than SOT in traditional STT and SOT of FM~\cite{STT vs SOT}, our results may imply that SOT in such AFMs has a great potential to realize even lower current density switching.

In contrast to the well-studied FM STT, while there are several reports on the electrical manipulation of AFM moments, there is no systematic study demonstrated to date.
This is because it is difficult to read out AFM moments electrically.
As the exchange bias generally appears at the interface between FM and AFM regardless of material combination, our results offer further experiments and may provide a deeper understanding of the electrical manipulation of AFM moments.

\section{ACKNOWLEDGMENTS}
The authors gratefully acknowledge Dr. K. Zhao, Prof. P. Gegenwart and R. Miki for characterization of the M$A$N single-layer films and T. Yoshida for the characterization of MnN/CFN bilayers. 
This work was supported by the Japan Society for the Promotion of Science (KAKENHI Grant Nos. 19K15445, 17K17801, and 17K19054), the Hori Science and Arts Foundation, and Kyosho Hatta Foundation.

\end{document}